\documentclass[submission,copyright,creativecommons,noncommercial]{eptcs}
\providecommand{\event}{FVAV 2017} % Name of the event you are submitting to
\usepackage{breakurl}             % Not needed if you use pdflatex only.
\usepackage{underscore}           % Only needed if you use pdflatex.
\usepackage{graphicx}
\usepackage{wrapfig}

\def\sectionname{Section}
\def\figurename{Figure}

\title{Game Theory Models for the  Verification of the Collective Behaviour of Autonomous Cars}
\def\titlerunning{Game Theory Models for the  Verification of the Collective Behaviour of Autonomous Cars}

\def\authorrunning{L.Z. Varga
}

\author{L\'{a}szl\'{o} Z. Varga
\institute{Faculty of Informatics\\
ELTE E{\"{o}}tv{\"{o}}s Lor\'{a}nd University 
Budapest, Hungary}
\email{\url{http://people.inf.elte.hu/lzvarga}}
}

\begin{document}
\maketitle

\begin{abstract}
The collective of autonomous cars is expected to generate almost optimal traffic. In this position paper we discuss the multi-agent models and the verification results of the collective behaviour of autonomous cars. We argue that non-cooperative autonomous adaptation cannot guarantee optimal behaviour. The conjecture is that intention aware adaptation with a constraint on simultaneous decision making has the potential to avoid unwanted behaviour. The online routing game model is expected to be the basis to formally prove this conjecture. 
\end{abstract}

\section{Introduction}
\label{sec:intro}

Autonomous cars open new possibilities and offer several benefits. These benefits include: increased mobility of the elderly and disabled people; better utilisation of travel time; finding urban places faster; more efficient traffic flow; less congestion; increased fuel efficiency. The last three benefits imply that the collective behaviour of autonomous cars will be close to a kind of optimum on the collective level. We are going to discuss the formal verification of this promise. Because we the verification is aimed at proving the properties of the collective behaviour of decentralised autonomous entities, the formal proofs are somewhat different from classical formal verification methods.

Autonomous cars detect their environment using different sensors like radar, LIDAR, GPS, computer vision, digital map, real-time traffic information and shared information. The planning unit of the autonomous car merges and interprets this sensory information to determine the necessary control actions to navigate the car to its destination and to avoid obstacles. As long as we focus on a single autonomous car, we can say that the planning unit executes {\it centralised adaptation}, because there is only one actor that senses the environment and takes actions to adapt to the changing environment. However if the road network is populated by several autonomous cars, then the overall traffic will emerge as the result of the collective behaviour of several autonomous cars. If there are several actors that sense the environment and take autonomous actions, then it is {\it decentralized adaptation}. If autonomous cars are designed with only centralized adaptation in focus, then they may be able to avoid obstacles and navigate to their destination, but if they meet other autonomous cars, then their joint actions may generate unwanted behaviour in some situations.

The human driven vehicle population may also bring about unwanted behaviour sometimes, but the issue will be more critical in the case of autonomous cars, because of two major differences between human driven and autonomous cars. 
One difference is that human drivers may be psychologically influenced, while autonomous cars always make {\it rational decisions}. Human drivers may follow their habits, although these habits may not be optimal and these habits may be unwanted for the overall traffic. The psychologically influenced decision of human drivers may sometimes result in preferable altruistic behaviour, but sometimes it may result in unwanted panic-like behaviour. On the contrary, autonomous cars always follow their designed rational preferences. The other difference is that human drivers may not always be aware of the relevant information, while autonomous cars always make {\it informed decisions}. Although humans are better in many cognitive tasks than machines, the machines have wider sensory capabilities than humans. Machines can use telecommunication technologies to ''see'' beyond objects (e.g. the approaching car behind the corner) and to ''see'' much farther away (e.g. congestion along the planned route on the other side of the city). As more and more information services are deployed to provide real-time traffic information to traffic participants, autonomous cars will have real-time and more precise information than humans. On the other hand, informed decision raises the issue of security and dependence on  technology. Information might be provided by malicious sources, or the data may not be reliable or accurate. Up-to-date information is a critical issue according to a recent study \cite{forbesweb}, but we presume that autonomous cars will receive exact enough real-time data.

The verification of centralised adaptation is necessary to ensure that autonomous cars can safely move on the street. Verification of decentralised adaptation is complementary to the verification of centralised adaptation, and it makes sure that the collective of autonomous cars do not produce unwanted behaviour, like for example inefficient usage of the road infrastructure. 

In \sectionname~\ref{sec:basic} we overview the main non-cooperative game theory models of decentralised adaptation and we highlight the main conclusions of the verification results from these models. In \sectionname~\ref{sec:improved} we discuss methods that improve the properties of decentralised adaptive systems, we present their models, and we highlight the main conclusions of the verification results from these models. In \sectionname~\ref{sec:conclusion} we conclude the paper with the conjecture that unwanted behaviour in decentralised adaptation could be avoided with the help of intention aware predictions, however this needs further research, which is among the goals of the investigations within the EFOP-3.6.3-VEKOP-16 project in connection with the RECAR \cite{recarweb} project.

\section{Non-cooperative Decentralised Autonomous Adaptation}
\label{sec:basic}

The basic characteristic of decentralised autonomous adaptation is that there are several  autonomous actors, called agents, which make decisions on which action to perform. The execution of the action of an agent uses limited resources, which are shared by several agents. The more agents use a resource, the less the agents prefer to use that resource. 
This is usually modelled with a cost function of the resource. 
The result of the action of an agent depends not only on its own action, but also on the action of all the other agents in its environment, because other agents may decide autonomously to use the same resource. An agent may not know which actions the other agents intend to do, therefore an agent may be uncertain which is the best action to perform. 

The decentralised decision making of the autonomous agents is the subject of multi-agent research \cite{wooldridge}. The current belief \cite{rosenschein} is that the best model of multi-agent decision making is founded in game theory  \cite{shoham}. In accordance with game theory \cite{agt}, the agents prefer some states of the environment to other states of the environment, which is modelled with a {\it utility} function. If the agents do not cooperate, then the rational agent selects that action which has the highest value among the worst possible utility of the outcomes of its own action and the actions of other agents. 
The actions of the agents are in {\it equilibrium} if no agent can benefit by changing its action while the other agents keep their actions unchanged. The {\it efficiency} of the multi-agent system can be measured as a combination of the utilities of all of the agents. A simple efficiency measure is  the sum of the utilities of all the agents. The multi-agent system is {\it optimal} if the sum of the utilities is maximal. Non-cooperative decentralized autonomous decision making may not lead to optimal result. This inefficiency is measured with the {\it price of anarchy} which is the ratio between efficiency measure of the equilibrium and the optimum. 

Game theory verified that the price of anarchy has an upper limit in some routing problems. The routing problem is a network with source routing, where end users simultaneously choose a full route to their destination and the traffic is routed in a congestion sensitive manner.  If the cost functions are linear functions of the traffic flow, then the price of anarchy is at most $4 \div 3$ \cite{routinggame}, i.e. this is how bad the overall traffic is when decentralised autonomous decision making is applied by the traffic flow. 

\begin{figure}
%\vspace{2.5cm}
%\centerline{\includegraphics[height=3.3cm] {figures-eps/congestion-game}}  %
\centerline{\includegraphics[width=\textwidth] {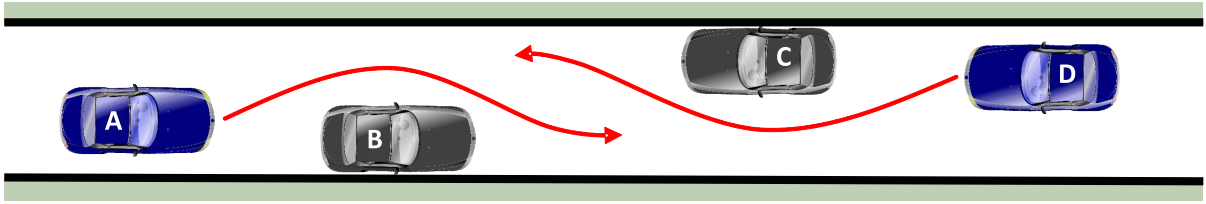}}  %
\caption{A simple minority game like situation}
\label{fig:minoritygame}
\end{figure}

There are games with many equilibria. In this case, if there is no coordination, then agents do not know which equilibrium is the goal of the collective and they may not select the right action. There are games where the equilibrium is not symmetric, i.e. some of the agents are not happy with the equilibrium, like in minority games \cite{Challet:2005}. In a minority game, the agents choose one of two choices independently, and the agents who end up on the minority side win. If every agent deterministically chooses the same action, then every agent is guaranteed to fail. The solution to this problem in game theory is to permit each agent to use a {\it mixed strategy}, where a choice is made with a particular probability. This may be good to model the macroscopic properties of diverse human behaviour, but random action {\it must not be allowed} to control autonomous cars in situations like for example on \figurename~\ref{fig:minoritygame}. Cars $B$ and $C$ are parked. The road is narrow and only one of the autonomous cars $A$ and $D$ can pass at a time. If both decide to go first (programmed randomly or deterministically), then they block each other.  If both decide to wait for the other, then the collective of cars $A$ and $D$ end up in a deadlock.

Classic game theory is concerned with the equilibrium, like traffic engineers who assume that the traffic is always assigned in accordance with the equilibrium  \cite{wardrop} \cite{beckmann}. They assume that all agents have complete knowledge about the game, and the agents come to the equilibrium with full rationality. This is not realistic, especially if agents do not cooperate and they have to make different actions depending on the actions of the others, because there are several equilibria. Therefore game theory investigated the {\it evolutionary dynamics} where the agents receive feedback by observing their own and other agents' actions and utility, and change their own actions based on these observations. It is verified that  with this feedback assumption, the above mentioned routing game converges to the equilibrium \cite{sandholm2001} \cite{fischer2004}. Another type of feedback is used in {\it regret minimisation}, where agents compare their actually experienced utility with the best possible utility in retrospect. It is verified that if the agents of the above routing game select actions to minimize their regret, then their behaviour will converge to the equilibrium  \cite{blum2006}. However the investigations of these game dynamics have the following assumptions: the decision making is on the flow level; the game is repeated; and the decision is based on experiences from the previous games. This is not realistic in autonomous cars, where the decision making is done at the individual car level, and the decision is based on the real-time situation instead of previous experiences. Nowadays more and more services provide real-time information about the overall traffic situation.

If we want to model that each subsequent agent of the traffic flow may select different route, depending on the current traffic situation, then the above game theory models cannot be used. Autonomous cars continuously enter the road network, and each agent of the traffic flow decides its optimal route when it enters the road network, and the decision is based on the real-time information about the status of the road network. The outcome travel time for a given agent depends not only on the current characteristic of the network and the route choice of all the agents simultaneously entering the road network, but also on the trip schedule of other agents that have entered the network previously, enter the network simultaneously, or will enter the network later.

The {\it online routing game} model (developed in \cite{Varga2014a}, and later refined in \cite{Varga2015a}) models this case when each autonomous car may select different route, based on real-time traffic information. In order to measure the efficiency of real-time data usage, the {\it benefit of online real-time data} concept was defined in \cite{Varga2014a}. The benefit of online real-time data is the ratio between the travel time with real-time data based planning and the travel time without real-time data based planning. The agents are happy with real-time data, if the benefit value is below $1$. Three types (worst/average/best) of benefit of online real-time data are needed in case an equilibrium traffic distribution cannot be achieved. It is proved in \cite{Varga2014a}, that if the agents try to maximise their utility, then the following properties are true: equilibrium is not guaranteed; ''single flow intensification'' is possible; and the worst case benefit value of online real-time data is not guaranteed to be below 1. Equilibrium may not be reached, because the traffic may fluctuate. ''Single flow intensification'' happens when vehicles entering the road network later may select alternative faster routes, and they may catch up with the vehicles already on the road, and this way they cause congestion. All-in-all, sometimes the traffic may produce strange behaviour \cite{Varga2015} and the collective of agents may be worse off by exploiting real-time information than without exploiting real-time information. 

The above results from the verification with the online routing game model indicates, that equilibrium cannot be verified if the agents autonomously follow their preferences to adapt to their environment. This is in line with experiments as well. For example the media supplement of \cite{nakayama} demonstrates that a small disturbance may bring about the fluctuation of the traffic and serious traffic jams are formed, if the agents apply non-cooperative decentralised adaptation. A specific algorithm can eliminate this effect if \href{https://d267cvn3rvuq91.cloudfront.net/v/files/experiment-a-plain-no-bookends.mp4?sw=600}{an autonomous car can proactively} force speed on others \cite{stern2017}. However in this experiment the autonomous car plays the role of a kind of central controller. 

We can conclude from the above models, proofs and verification results, that non-cooperative decentralised autonomous adaptation to real-time data cannot guarantee to avoid unwanted behaviours of the collective of autonomous cars. The optimal traffic flow is not guaranteed, and  an equilibrium traffic flow, which is worse than the optimal, is not guaranteed either.

\section{Improved Decentralised Autonomous Adaptation}
\label{sec:improved}

Some form of cooperation should be built into autonomous cars to improve their collective behaviour. Cooperation is not only information exchange, but also coordinated actions within an agent community, which means that an agent's behaviour is influenced by the intentions and results of other agents. Cooperation proved to be useful in other application areas as well \cite{Cockburn1992}.

There are several ways to coordinate the actions of agents. One way is to centralise the multi-agent system and assign a {\it control authority} above the agents. For autonomous cars this would mean for example that each geographical territory would be under the control of a control authority. When an autonomous car reaches such territory, then it checks in at the control authority, and after the check-in, the control authority would tell the autonomous car the exact route to follow to its destination.  This is somewhat similar to the operation of airports and how airplanes move in the area of the airport. The control authority is centralised, therefore it can be verified with verification methods of centralised systems. In this control authority approach the autonomous car becomes something similar to a remote controlled car. The users of autonomous cars may not accept this concept, especially if the commands of the central authority are not in line with the personal preferences and the individual has to suffer major drawback to facilitate the collective benefit. The central control authority is a reliability risk and a performance bottleneck as well.

\begin{wrapfigure} {r}{0.35\textwidth}
%\vspace{2.5cm}
%\centerline{\includegraphics[height=3.3cm] {figures-eps/congestion-game}}  %
\centerline{\includegraphics[width=0.3\textwidth] {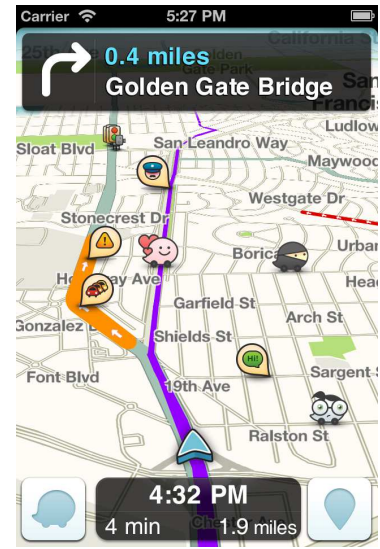}}  %
\caption{A service from the Google Play Store with prediction potential}
\label{fig:predictionservice}
\end{wrapfigure}

If the idea of the central control authority is given up, then another approach with some autonomy is when the coordination and communication is fostered by some kind of central service, but the control remains at the agents. The agents do not communicate directly with each other, but they communicate their intentions with the central service. The central service aggregates data about the agent collective and sends feedback to the agents \cite{claes2014}. The {\it intention aware} \cite{weerdt2016} and {\it intention propagation} \cite{claes2011} approaches are based on this scheme.  When an agent has made a decision on its planned route, then it sends its selected intention to the central service. The central service is able to make a forecast of the future traffic situation based on the current traffic state and the communicated intentions of the agents. The central service provides the traffic forecast back to those agents who are still planning their trips, and these agents use this information to plan their trip, and when they have made a decision, then they also communicate their intentions to the central service. In theory, the online navigation software like \href{http://maps.google.com/}{Google Maps} and \href{http://waze.com}{Waze} (\figurename~\ref{fig:predictionservice}) know the intentions of the agents and could use this information to make predictions.

The online routing game model was used to formally verify if the prediction power of the central service can solve the problem of avoiding unwanted behaviours of the collective of autonomous cars. The verification results \cite{Varga2015a} show that there is no guarantee on the value of the worst case benefit of online data and there is no guarantee on the equilibrium, i.e. the traffic may fluctuate. This means that if the agents selfishly exploit intention aware prediction, then in some networks and in some cases the traffic may be worse off by exploiting real-time information and prediction than without. This is due to the fact that the central service cannot take into account the decisions of those agents that make decisions simultaneously, and the decisions may depend on each other. However, it is verified \cite{Varga2016a} that in a small but complex enough network, where there is only one traffic flow and therefore the agents that follow each other do not make decisions at the same time, then there is a guarantee on the value of the worst case benefit of online real-time data with prediction. In this case the agents might just slightly be worse off with real-time data and prediction in some cases. This verification result shows, that in the network of \cite{Varga2016a}, the intention aware prediction establishes enough coordination among the agents. Currently the conjecture is that if simultaneous decision making among the agents is prevented, then intention aware prediction can limit the fluctuation in the multi-agent system \cite{Varga2016} and the traffic converges to the equilibrium in bigger networks as well. This conjecture is an important challenge for the verification of the collective behaviour of autonomous cars.

A critical issue of intention aware prediction is the trust in the intention submissions. If agents can profit from misleading other agents with revealing false intention, then they might be tempted to exploit this. In case of autonomous cars, the intention submission is done by the software built into the car. Car manufacturers will have to certify that their software submits its intention truthfully and correctly, and the software cannot be modified.

If the idea of a central coordination service is given up, then the agents have to coordinate their activities on a peer-to-peer basis. Two possible approaches are coalition formation and gossiping. Coalition formation may improve the behaviour of the agent collective, however if an agent can benefit from breaking the coalition agreement, then an authority is needed to make sure that the coalition agreement is kept by the agents. The modelling and verification of coalition formation of autonomous cars can be founded in cooperative game theory \cite{chalkiadakis2011}. Gossiping may be a means for spreading information \cite{demers1987} or aggregating information \cite{kempe2003}, however it is up to the agents how they use this information. The most likely usage is to predict future traffic situation as in the case of intention awareness, but gossiping does not need a central service. As we have seen, the guaranteed benefits of the prediction of future traffic situation has not yet been verified.

We can conclude from the above discussion, that cooperation techniques based on intention awareness have been proposed, but the verification of the preferred collective behaviour of the agents is still a challenge.

\section{Conclusion}
\label{sec:conclusion}

One of the main promises of autonomous cars is that they produce better and closer to optimum collective behaviour than human drivers do. Each member agent of the collective of autonomous cars tries to adapt to the changing environment, therefore they execute decentralised  autonomous adaptation. The verification of the preferred decentralised autonomous adaptation is important, because the collective of agents may produce unwanted behaviour, especially if they exploit real-time information about the whole collective. We have discussed models of collective agent behaviour, and discussed the verification results from these models.  In this position paper, we argued that guaranteeing the avoidance of unwanted collective behaviour of non-cooperative agent collectives cannot be verified. Improved decentralised autonomous adaptation techniques try to establish some kind of cooperation among the agents, mainly through intention awareness. We have discussed models and verification results of intention aware collective agent behaviour. The verification process started recently, and the verification results from these models show that intention awareness improves the collective behaviour, but simultaneous decision making may still cause problems. The aggregation of intentions to predict future traffic state needs further research, too. We argued that these are critical issues, because if the wanted behaviour cannot be verified, than the only viable approach to ensure the close to optimum behaviour of the collective of autonomous agents is centralised control.  We aim to discuss this at the workshop and investigate this issue in a future  project.

\nocite{*}
\bibliographystyle{eptcs}
\bibliography{DAS}

\end{document}